\documentclass[aps,prb,twocolumn, superscriptaddress, showpacs]{revtex4}
\usepackage{amsmath}
\usepackage{amssymb,amsfonts}
\usepackage{graphicx}
\usepackage{lmodern}
\usepackage{slantsc}
\usepackage{scalefnt}
\usepackage{subfigure}
\usepackage{dsfont}
\usepackage{xspace} 
\usepackage{ifpdf}
\usepackage{tensor}
\usepackage{array}
\allowdisplaybreaks[1]
\usepackage[mathcal]{euscript}
\usepackage{pbox}

\newcommand{\be}{\begin{equation}}
\newcommand{\ee}{\end{equation}}
\newcommand{\bse}{\begin{subequations}}
\newcommand{\ese}{\end{subequations}}

\newcommand{\Z}{\mathbb{Z}}
\newcommand{\ii}{\mathrm{i}}
\newcommand{\e}{\mathrm{e}}

\newcommand{\N}{\mathcal{N}}

\newcommand{\T}{\mathcal{T}}
\newcommand{\A}{\mathcal{A}}

\newcommand{\U}{\mathcal{U}}
\newcommand{\C}{\mathcal{C}}
\newcommand{\m}{\mathbf{m}}
\renewcommand{\b}{\mathbf{b}}
\newcommand{\bpm}{\begin{pmatrix}}
\newcommand{\epm}{\end{pmatrix}}
\newcommand{\bmm}{\begin{matrix}}
\newcommand{\emm}{\end{matrix}}

\newcommand{\x}{\times}
\newcommand{\ox}{\otimes}
\newcommand{\II}{$\text{Ising}\times\overline{\text{Ising}}$\xspace}

\newcommand{\diag}{\mathrm{Diag}}

\newcommand{\Hom}{\mathrm{Hom}}
\newcommand{\id}{\mathrm{id}}
\newcommand{\rep}{\mathrm{Rep}}

\newenvironment{smallarray}[1]
 {\null\,\vcenter\bgroup\scriptsize
  \arraycolsep=.13885em
  \hbox\bgroup$\array{@{}#1@{}}}
 {\endarray$\egroup\egroup\,\null}

\newcommand{\sixj}[6]{
\bigl\{\hspace{-0.2em}
\begin{smallarray}{cc|c}
    #1 & #2 & #3\\ [0.2em]
    #4 & #5 & #6
\end{smallarray}\hspace{-0.2em}\bigr\} }

\newcommand{\Fm}[7][0]{
\ifthenelse{\NOT \equal{#1}{0}}{
\left[F^{#2#3#4}_{#5}\right]_{#6#7}
}
{\bigl[F^{#2#3}_{#4#5}\bigr]^{#6}_{#7}}
}

\makeatletter
\newcommand*{\Relbarfill@}{\arrowfill@\Relbar\Relbar\Relbar}
\newcommand*{\xeq}[2][]{\ext@arrow 0055\Relbarfill@{#1}{#2}}
\makeatother

\begin{document}

\title{Generalized ADE Classification of Gapped Domain Walls}

\author{Ling-Yan Hung}
\email{jhung@perimeterinstitute.ca}
\affiliation{Department of Physics and Center for Field Theory and Particle Physics, Fudan University, Shanghai 200433, China}
\author{Yidun Wan}
\email{ywan@perimeterinstitute.ca}
\affiliation{Perimeter Institute for Theoretical Physics, Waterloo, ON N2L 2Y5, Canada}

\date{\today}

\begin{abstract}
In this paper we would like to demonstrate how the known rules of anyon condensation motivated physically proposed by Bais \textit{et al} can be recovered by the mathematics of twist-free commutative separable Frobenius algebra (CSFA). In some simple cases, those physical rules are also sufficient conditions defining a twist-free CSFA. This allows us to make use of the generalized $ADE$ classification of CSFA's  and modular invariants to classify anyon condensation, and thus characterizing all gapped domain walls and gapped boundaries of a large class of topological orders. In fact, this classification is equivalent to the classification we proposed in Ref.\onlinecite{HungWan2014}. \end{abstract}
\maketitle

\section{Introduction}\label{sec:intro}
Classification of two-dimensional matter systems with intrinsic topological order can deepen our understanding of their relations with topological field theories, conformal field theories, and anomalies in field theories, and will thus boost their theoretical and experimental applications. For the latest perspective regarding these connections and summary of results, see Nayak\cite{Nayak2008} and Wen and Kong\cite{Kong2014}, and the references therein. It is believed that two-dimensional bosonic topological orders are described by unitary modular tensor categories (UMTCs)\cite{Kitaev2006,Rowell2009}.  For each given phase, it is important to understand the properties of their boundaries and defects, specifically,  gapped domain-walls (GDWs) separating different phases. Even within a single phase, there could potentially be \textit{transparent} GDWs which are closely related to global symmetries of the topological phase. If one of the two phases involved is the vacuum, the GDW reduces to a gapped boundary (GB). A systematic understanding and classification of GDWs therefore supplies extra physical information about a given topological phase, and leads to a web of connections between phases. Ref.\onlinecite{Wang2012,Barkeshli,Barkeshli2013c} have offered classifications of GBs of Abelian topological phases. Recently, the idea of anyon condensation due to Bais \textit{et al}\cite{Bais2002,Bais2009,Bais2009a} has been applied to studying the GDWs between two topological phases\cite{Bais2009,Kitaev2012,Kong2013}, and from which one could compute the ground state degeneracy of phases with boundaries\cite{HungWan2014,Lan2014}. Moreover it is known that classifying anyon condensation is also connected to classifying symmetry-enriched topological phases.\cite{Hung2013,Barkeshli2014c}  To make progress towards a systematic classification, it is important to unravel how the physics of anyon condensation is connected to the mathematics. With such a goal in mind, we are reporting various interesting progress: 
\begin{enumerate}
\item We demonstrate explicitly that Bais-Slingerland rules of anyon condensation are implied mathematically from the concept of twist-free CSFA in a UMTC. This gives further support of the correspondence between these concepts: anyon condensation, quantum subgroup, vertex operator algebra embedding, modular invariants.
and GDWs.%
\item We then work backwards, showing that for a large class of anyon condensation---simple currents condensation---and for electric and magnetic condensation in the quantum double of any finite gauge group $G$ following from Bais' and Slingerland's heuristic rules \textbf{is in 1-1 correspondence with a modular invariant. It therefore defines a twist-free CSFA.}%
\item These imply that anyon condensation, and thus GDWs, is classified by the mathematics of twist-free CSFA. Hence, they are also classified by the generalized $ADE$ Dynkin diagrams, also known as fusion graphs. These are known to classify quantum subgroups and the modular invariants of rational conformal field theories (RCFTs)\cite{CFTbook,Kirillov2002,Fuchs2002}.

\item A transparent GDW is uniquely characterized by the mass matrix $\m$ of a modular invariant, whereas a generic GDW or a GB is uniquely characterized by the branching matrix $\b$ of a vertex operator algebra (VOA) embedding. 
%
%
\end{enumerate}

Explicit examples will be used to illustrate the results listed above. Note that in a recent work\cite{Lan2014}, GDWs are described by some matrix $W$ satisfying a set of constraint equations. Since the mass matrices and branching matrices $\m$ and $\b$ matrices\cite{Fuchs2002} and the $W$ matrices are both describing modular invariants and in fact agree in every known example, our study suggests that they are likely equivalent mathematically. The techniques developed in the mathematics literature is promising to give us a systematic classification of the $W$ matrices.

To keep the physics as clear as possible in the main part of the paper, we present in the appendix a review of Bais-Slingerland rules, a glossary for twist-free CSFA, $su(2)_k$ modular data, and how we use the modular invariants of $su(2)_k$ to classify the GDWs and GBs in the corresponding topological phases. Also in the appendix, we record our finding of some properties and implications of the $W$ matrices. Note that throughout this paper, we may abuse the language by referring to a topological phase as a UMTC, or in the cases applicable as a quantum group or an affine Lie algebra at certain level whose representation category is a UTMC describing the topological phase. We may also loosely use anyons, (topological) sectors, and (primary) fields interchangeably.

\section{Bais-Slingerland anyon condensation $=$ twist-free CSFA }\label{sec:Bais}
Bais \textit{et al} have developed a set of consistent rules for anyon condensation that can break a topological phase, described by a UMTC $\C$, into a smaller (i.e., with fewer anyon types and smaller total quantum dimension) topological phase described by a UMTC $\U$. This mechanism, though has not seen any counter-examples, is based on a set of \textit{ad hoc} rules. In this section, we shall give these rules a mathematical foundation by showing their equivalence to what is known as a twist-free CSFA and the mathematical structures induced by this algebra. We tabulate this equivalence as follows.
\begin{widetext}

\begin{table}[h!]
\begin{tabular}{c||c}
\toprule
Parent topological phase $\C$ & MTC $\C$ \\\hline
\pbox{10cm}{Bais' condensable set of anyons \\ $A=\{\gamma_i|i=0,\dots,m,m<|\C|,\gamma_0:=1_\C\}$} & twist-free CSFA $A=\oplus_{i=0}^m\gamma_i\in \C$ \\\hline
$\theta_{\gamma_i}=1$ & $\theta_A=\id_A$\\\hline
Intermediate phase $\T$ & $\rep A$, a tensor category \\\hline
The condensates constitute the vacuum $1_\T\in \T$ & $A=1_{\rep A}$\\\hline
\pbox{10cm}{Anyon $a$ may split into $p$ parts in $\T$: $a\rightarrow \sum_{j=1}^p a_j$ } & \pbox{10cm}{Object $a\otimes A\big|_{\rep A}=\oplus_{j=1}^p X_j$, $X_j\in\rep A$ is simple}  \\\hline
\pbox{10cm}{Conservation of quantum dimension: $d_a=\sum\limits_{j=1}^p d_{a_j}$} & \pbox{10cm}{$\dim_\C a = \frac{\dim_\C a\ox A}{\dim A} = \sum\limits_{j=1}^p \frac{\dim_\C X_j}{\sum\limits_{i=0}^m\dim_{\gamma_i}}=\sum\limits_{j=1}^p\dim_A X_j$} \\\hline
Unconfined phase $\U$ & $\rep_0A$, a braided tensor category\\\hline 
$\U$ contain the $\T$ anyons that are mutually local with $A$ & $\rep_0 A=\{X\in\T|(A\otimes_A X)R_{XA}R_{AX}=A\otimes_A X\}$\\\hline
\end{tabular}
\caption{Correspondence between Bais-Slingerland anyon condensation and twist-free CSFA.}
\label{tab:equiv}
\end{table}
\end{widetext}

Table \ref{tab:equiv} exhibits the correspondence between Bais-Slingerland anyon condensation and twist-free CSFA. The table is self-explanatory, supplemented by the glossaries we provide in the appendix. In what follows, we would like to address the most crucial points in this correspondence, with each point illustrated by an explicit example: the UMTC as the representation category of the affine Lie algebra $su(2)_{10}$ that is isomorphic to the quantum group $\U_q(su(2))$ with $q=\exp(\ii\pi/6)$.

\textit{What condenses}? In Bais-Slingerland condensation, the condensable anyons are self-bosons. In a topological phase $\C$, such that the entire set $A$ of condensable anyons can condense together, this set should be be closed under fusion, in the sense that the fusion of any two anyons in $A$ must contain at least one anyon in $A$, and any anyon in $A$ must appear in the fusion product of two anyons in $A$. In fact, as can be easily shown, these two conditions have a physical consequence:  any two anyons in $A$, say $a$ and $b$, have trivial monodromy with respect to at least one fusion channel. Namely, $\exists c\in a\otimes b$, such that, $M^c_{ab}=\theta_c/(\theta_a\theta_b)=1$. This defines the notion that the anyons in $A$ are mutually local with respect to each other. Since $A$ condenses to be the new vacuum, its self-monodromy and self-fusion must commute. Consider the example of $su(2)_{10}$. This topological phase has $11$ elementary anyons, labeled by integers from $0$ to $10$, where $1_\C:=0$. The topological properties of these anyons are listed in Table \ref{tab:su2_10} in the appendix. Clearly, the anyon $6$ with $h_6=1$ is the only self-boson in the spectrum and meets the criteria of Bais-Slingerland condensation. 

The above properties of a condensable set $A$ matches precisely the defining properties of a twist-free CSFA. First, a  CSFA $A$ in a UTMC $\C$ is an object in $\C$. This object is generally nonsimple and take the form $A=\oplus_i\gamma_i$, where $\gamma_i$'s are simple objects of $\C$. Such an algebra $A$ must be closed under fusion, which means there exists a product (a projection): $A\otimes A\rightarrow A$. This product is associative, and the commutativity requires this product to commute with the self monodromy of $A$. Second, twist-free means $\theta_A=\id_A$, which implies $\theta_{\gamma_i}=1$, for all $\gamma_i$ appearing in $A=\oplus_i\gamma_i$. Third, a CSFA is self-dual\cite{Kirillov2002}. Therefore, the condensable set $A$ in a topological phase described by a UMTC $\C$ indeed comprises a twist-free CSFA.
Again in the example of $su(2)_{10}$, we have $A=0\oplus 6$, which can be easily checked to be a twist-free CSFA. The separability of a CSFA will be useful shortly in the following.

\textit{Conservation of quantum dimension}. As explained in the appendix, a twist-free CSFA of $A\in\C$ induces another category $\rep A$, the category of modules over the algebra $A$. The separability of the CSFA $A$ ensures that $\rep A$ is semisimple, admitting the notions of simple objects and non-simple objects as direct sums of simple objects. The commutativity of $A$ guarantees that $\rep A$ is a tensor category. The splitting of an anyon in a condensation $A$ can be understood in the categorical language as follows. Since $A$ becomes the unit object, i.e., the vacuum in $\rep A$, an object in $\rep A$ is an object in $\C$ equipped with an action by $A$. A convenient way of studying the objects in $\rep\ A$ is via a map (a functor) $F: \C\rightarrow\rep A$, such that for a $V\in\C$, $F(V)=A\ox V\in\rep A$. Note that $V$ is not necessarily simple in $\C$, and $A\ox V$ not necessarily simple in $\rep A$. More interestingly, even if $V$ is a simple object in $\C$, i.e., an elementary anyon, let us rename it as $a:=V$, $a\ox A$ may still be non-simple in $\rep A$. This happens typically when $\dim_a\geq 2$: for such a simple $\C$ object $a$, if it appears $p$ times in $a\ox A$, $a\otimes A$ will be a direct sum of $p$ simple objects in $\rep A$ if all multiplicities of the splittings $b_{i a}$ are unity.  Namely, $a\otimes A|_{\rep A}=\oplus_{j=1}^p X_j$, where $X_j$'s are simple in $\rep A$. Keep in mind that each $X_j$ here may appear in $\C$ as a non-simple object. More generically, we have \be p = \sum_j (b_{j a})^2\ee.  One can see that such a decomposition of $a\ox A$ in $\rep A$ corresponds precisely to the notion of a $\C$ anyon splitting into $\T$ anyons according to Bais-Slingerland rules of anyon condensation (see appendix). This kind of decomposition also manifests the conservation of quantum dimension in Bais-Slingerland rules, as presented in the $7$-th row of Table \ref{tab:equiv}. 

Let us go back to the example of $su(2)_{10}$. This case has quite a few occurrences of splitting but we focus on only two of them to illustrate the point. The dimension of a CSFA $A$ is by definition $\dim A=\sum_{i=1}^m d_{\gamma_i}$, so we have in the current example $\dim A=d_0+d_6=3+\sqrt{3}$. This quantity bears the name quantum embedding index in Bais.\cite{Bais2009a} According to Bais-Slingerland rules, the fusion $6\ox 6=0\oplus 2\oplus 4\oplus 6\oplus 8$ would require the splitting $6\rightarrow 6_1+6_2$, with $d_{6_1}=1$ and $d_{6_2}=1+\sqrt{3}$. On the other hand, the fusion rule $3\ox 6=3+5+7+9$ implies $3\rightarrow 3_1+3_2$ with $d_{3_1}=\sqrt{2+\sqrt{3}}$ and $d_{3_2}=\sqrt{2}$.
On the side of the algebra $A$, the quantum dimension of a $\rep A$ object in the form of $a\ox A$ reads $\dim_A(a\ox A):=\dim_\C(a\ox A)/\dim A=\dim_\C(a)\dim A/\dim A=\dim_\C a$.  If $\dim_\C a<2$, $a\otimes A$ must be a simple object in $\rep A$. In the current case, e.g., $F(2)=2\oplus 4\oplus 6\oplus 8$ and $F(9)=3\oplus 5\oplus 9$ are simple in $\rep A$. In the language of twist-free CSFA, the splitting of $6$ is understood as $F(6)=6\ox A=A\oplus( 2\oplus 4\oplus 6\oplus 8)=A\oplus F(2)$, and that of $3$ is $F(3)=F(9)+(3\oplus 7)$, where $3\oplus 7$ is also simple in $\rep A$ but not of the form of $F(V)$ for any $V\in su(2)_{10}$. We can verify that $\dim_A(3\oplus 7)=(d_3+d_7)/\dim A=\sqrt{2}$. Then one can explicitly check that $\dim_A F(6)=d_6=\dim A+\dim_A F(2)= 1+\sqrt{3}$, as well as $\dim_A F(3)=\dim_AF(9)+\dim_A(3\oplus 7)=d_9+\sqrt{2}=\sqrt{2+\sqrt{3}}+\sqrt{2}$. As such we can make the identifications $A=6_1$, $F(2)=6_2$, $F(9)=3_1$, and $3\oplus 7=3_2$. The identification $A=6_1$ corroborates the argument according to Bais-Slingerland rules that the condensate $6$ does not completely condense but splits into two portions, only one of which actually condenses. 

According to Bais-Slingerland rules, the fusion rules of $\rep A$ commute with the splitting. This follows automatically from the tensorial property of the map $F$, i.e., $F(V\otimes W)=F(V) \otimes_A F(W)$, where $\otimes_A$ denotes the fusion in $\rep A$.\cite{Kirillov2002}

A fact is that although the UMTC $\C$ we begin with is a braided tensor category, the $A$-induced category $\rep A$ is in general not braided. To obtain a braided tensor category from $\rep A$, one would have to exclude those $\rep A$ objects that are mutually nonlocal with respect to $A$. This procedure is showed in the last row of Table \ref{tab:equiv}. The so obtained braided tensor category is denoted $\rep_0 A$\cite{Kirillov2002,Frohlich2006}. If the CSFA $A\in \C$ under consideration is twist-free, then $\rep_0 A$ is also a UMTC. It is straightforward to see that such a $\rep_0 A$ is an unconfined phase $\U$ in the context of Bais; however, for the converse statement, we are not able to prove it in general rigorously except for the cases with simple-current condensation in chiral topological phases and  electric/magnetic condensation in a quantum double of a finite gauge group $G$. The proof for these cases will be demonstrated in the next section. 

Back to the example of $su(2)_{10}$, the $\rep_0 A$ contains three simple objects, $1:=A|_{\rep A}$, $\sigma:=(3\oplus 7)_{\rep A}$, and $\psi= (4\oplus 10)_{\rep A}$, with $d_1=1$, $d_\sigma=\sqrt{2}$, and $d_\psi=1$. They satisfy Ising type fusion rules. As a UMTC, in this case, $\rep_0 A=so(5)_1$.
It is a theorem\cite{Kirillov2002} that the total quantum dimensions of $\C$ and $\rep_0 A$ are related by
\be\label{eq:qdrel}
\dim\rep_0A=\frac{\dim\C}{\dim A}.
\ee
This formula is easily checked in the current example.

\begin{figure}[ht!]
\centering
\includegraphics[scale=1]{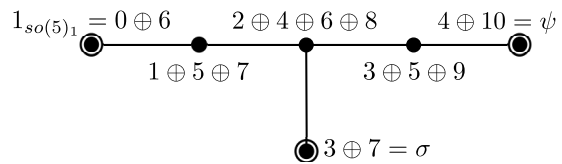}
\caption{The $E_6$ Dynkin diagram for the embedding $su(2)_{10}\subset so(5)_1$. The nodes are labeled by the simple objects in $\rep A$, whereas the three end-nodes are the simple objects in $\rep_0 A=so(5)_1$. A similar figure also appears in Ref.\onlinecite{Kirillov2002}.}
\label{fig:E6dynkin}
\end{figure}

In this example, one sees an interesting mathematical structure, namely, the embedding $su(2)_{10}\subset so(5)_1$, which is an instance of the vertex algebra embedding in RCFTs. This also corresponds to the quantum subgroup relation $\U_{\exp{\ii\pi/6}}(su(2))\supset\U_{\exp\ii\pi/2}(so(5))$. This relation is captured precisely by the Dynkin diagram of the $E_6$ Lie algebra, shown in Fig. \ref{fig:E6dynkin}.
This figure is self-explaining. The simple objects in $\rep A$ correspond to the simple roots of the $E_6$ Lie algebra, whereas the simple ones in $\rep_0 A$ label those simple roots at the ends of the three legs of the diagram. The tensor product between the $E_6$ simple roots describe the fusion rules of the $\rep A$ simple objects. This diagram also characterizes a nondiagonal modular invariant of the $su(2)_{10}$ RCFT, namely, 
\be\label{eq:E6inv}
Z^{su(2)_{10}}_{E_6}=|\chi_0+\chi_6|^2+|\chi_3+\chi_7|^2+|\chi_4+\chi_{10}|^2,
\ee 
which is called the $E_6$ invariant of $su(2)_{10}$.
One sees that the confined objects due to the condensation $A=0\oplus 6$ are absent from the modular invariant \eqref{eq:E6inv}. The relation between the $su(2)_{10}$ anyons and the $so(5)_1$ ones is exactly captured by what is known as a branching matrix $\b_{\alpha a}$ that maps the $su(2)_{10}$ anyons $a$ to the $so(5)_1$ anyons $\alpha$, which reads
\be
\b=\bordermatrix{
   & 0 & 1 & 2 & 3 & 4 & 5 & 6 & 7 & 8 & 9 & 10\cr
 1 & 1 & 0 & 0 & 0 & 0 & 0 & 1 & 0 & 0 & 0 & 0 \cr
\sigma & 0 & 0 & 0 & 1 & 0 & 0 & 0 & 1 & 0 & 0 & 0 \cr
\psi & 0 & 0 & 0 & 0 & 1 & 0 & 0 & 0 & 0 & 0 & 1
}.\label{eq:su2_10bm}
\ee  
In general for an anyon condensation $A$ that corresponds to a vertex operator algebra embedding or quantum subgroup structure, we have
\be\label{eq:bmatrix}
\alpha=\oplus_a\b_{\alpha a}a,\quad a\in \C, \alpha\in\rep_0 A,\b_{\alpha a}\in\Z_{\geq 0}.
\ee
One can view that the gapped domain wall between the topological phases $su(2)_{10}$ and $so(5)_1$ is characterized by either the $E_6$ Dynkin diagram in Fig. \ref{fig:E6dynkin} or by the branching matrix \eqref{eq:su2_10bm}.
The branching matrix \eqref{eq:bmatrix} commutes with the modular $S$ and $T$ matrices of the parent and child topological phases in the following sense.\cite{Fuchs2002,Gaiotto2012}
\be\label{eq:STbST}
\begin{aligned}
&\sum_\beta S_{\alpha\beta}\b_{\beta a}=\sum_{b} \b_{\alpha b}S_{ba},\\
&\sum_\beta T_{\alpha\beta}\b_{\beta a}=\sum_{b} \b_{\alpha b}T_{ba},
\end{aligned}
\ee
where $(T)S_{\alpha\beta}$ and $(T)S_{ab}$ are respectively the $(T)S$ matrices of $\rep_0 A$ and $\C$.
This is reasonable because the modular invariant induced by an anyon condensation $A$ is indeed a modular invariant partition function of the RCFT corresponding to the child phase.


In fact, anyon condensation in chiral topological phases described by representation categories of $su(2)_k$, or equivalently of $\U_{\exp\ii 2\pi/(k+2)}(su(2))$ is fully classified by the Dynkin diagrams of the $ADE$ type. For anyon condensation $A$ in more general affine Lie algebra or the equivalent quantum groups, the fusion rules of $\rep A$ simple objects give rise to fusion graphs, which are referred to as generalized $ADE$ Dynkin diagrams. Only $ADE$ type can arise because the fusion graphs are always simply-laced for twist-free CSFAs.\cite{Kirillov2002} A list of this classification for all affine Lie algebras up to certain ranks and high levels can be found in Ref.\onlinecite{Gannon2005}. Therefore,  we can claim that the gapped domain walls between chiral topological phases that fall into this large family are fully classified by the generalized $ADE$ Dynkin Diagrams, as well as by the branching matrices of the corresponding vertex operator algebra embedding.

One may notice that a parent phase always corresponds to a VOA smaller than what its child phase corresponds to. The quick physical reason is, the condensation in the parent phase reduces the spectrum or the Hilbert space of the parent topological phase by forcing the condensates to be part of the new vacuum; however, on the corresponding RCFT side, this would lead to more conserved currents that enlarge the symmetry, and hence result in a larger VOA. A more mathematical account for this relation can be found in Ref.\onlinecite{Bais2009a}. 

Recognizing the correspondence between anyon condensation and twist-free CSFAs also hints at a more convenient way of finding the $\rep A$ simple objects and their fusion rules for a condensation $A$. We present our simpler rules as follows.
\begin{enumerate}
\item For any $a\in\C$ whose $d_a<2$, $a\ox A$ must be simple in $\rep A$. 
\item For any $a\in\C$, if $a$ appears $p\geq 2$ times in $a\otimes A$, $a\otimes A$ must split into $p$ simple objects in $\rep A$, each of which contains a copy of $a$ in the case where all $b_{i a}=1$. 
\end{enumerate}

The two rules above combined can lead us to all the simple objects in $\rep A$ and can be automated by programming. As long as all the simple objects in $\rep A$ are found, their fusion rules can be nailed down easily. These rules are useful when we are dealing with a topological phase that is not listed in any table of classification yet. Of particular interest are non-chiral topological phases. One large family of non-chiral topological phases consists of the (twisted) quantum double of a finite group, which will be discussed  in the next section. Another large family is comprised of the doubles of chiral topological phases described by quantum groups, namely $\U_q(\mathfrak{g})\x\overline{\U_q(\mathfrak{g})}$.

A doubled phase $\C$ usually has a lot more condensable anyons than its chiral component and thus more than one twist-free CSFAs, i.e., more than one set of anyon condensation. Typically, all the diagonal pairs of the anyons in the two chiral components are self-bosons. An interesting subtlety often arises in a doubled phase. Namely, two anyon condensations $A$ and $A'$ maybe Morita-equivalent, $A\sim A'$, in the  sense that they induce equivalence $\rep_0 A\sim \rep_0 A'$. We leave the definition of Morita-equivalence to the references (see e.g. \cite{Kitaev2012}). A quick example lies in the case with $su(2)_{10} \x \overline{su(2)_{10}}$, where $A=0\bar 0\oplus 6\bar 0$ and $A'=0\bar 0\oplus 0\bar 6$. Another example, the simplest one, is the $\Z_2$ toric code phase, where the two condensations $A=1\oplus e$ and $A'=1\oplus m$ are Morita equivalent as they break the topological phase to the trivial phase.

The $ADE$ classification not only includes the GDWs between two distinct topological phases but also encompasses those GDWs between two copies of the same topological phase. On the other hand, the $ADE$ classification not only works in the chiral case but also handles the nonchiral doubling of the chiral phases. But we shall leave the discussion of these points and the subtleties therein to Sec. IV.

\section{Simple currents condensation vs Modular Invariance}\label{CFT}

We would like to demonstrate here that using Bais-Slingerland rules, simple currents condensation obtained are in 1-1 correspondence with known modular invariants. To do so, we need to collect various facts in the dual RCFT of the topological phase. In an RCFT, simple currents are primaries that have unique fusion product with any primary in the spectrum of the RCFT.  Most importantly, the simple currents in an RCFT all have quantum dimension unity and form a cyclic group $Z_N$ under fusions, with $N$ being the order of the simple current $\varphi_0$, such that all simple currents in the theory are merely fusion powers of $\varphi_0$ up to order $N$. It can be shown\cite{Fuchs1992} that the conformal dimension $h_m$ of the simple current $\varphi_0^m$ takes the following form.
\be\label{eq:simpleCurrentCD}
h_m=\zeta\frac{m(N-m)}{2N}\pmod{\Z},
\ee
where $\zeta$ is an integer defined modulo $N$ for odd $N$ and otherwise modulo $2N$. These conformal dimensions have been computed for all affine Lie algebras that have simple currents\cite{Fuchs2002} but we do not need their explicit values for our purposes here. Worth of note is that for chiral CFTs, according to the formula above, a simple current may not have integer conformal dimension but those corresponding to anyon condensation should have integer conformal dimensions. The simple currents in a RCFT act like permutations on the spectrum of the theory, and thus divide the spectrum into different orbits by fusing with the simple currents. On such an orbit, we arbitrarily choose a primary and denote it by $\phi_i\equiv \phi_{i^0}$, then any other primary $\phi_{i^n}$ on the orbit is obtained by
\be\label{eq:orbit}
\phi_{i^n}=\phi_i\x \varphi_0^n,
\ee
where $n=1,\dots,N$.
Consequently, the conformal dimensions of $\phi_{i^n}$ and of $\phi_i$ are related by the formula\cite{SCHELLEKENS1990,Fuchs1992}.
\be\label{eq:orbitCurrentCD}
h_{\phi_{i^n}}=h_{\phi_i}+\zeta\frac{n(N-n)}{2N}-\eta\frac{n}{N}\pmod{\Z},
\ee   
for some integer $\eta$. Let us denote such an orbit by $\{\phi_i^n\}$.

The idea is to arrange some of the orbits of simple currents as blocks comprising the modular invariant. In an RCFT, if one simple current has integer conformal dimension, all simple currents generated should also have integer conformal dimensions. It is known that an RCFT that has simple currents with integer conformal dimensions can possess the follow integral spin invariants.
\be\label{eq:intSpinModInv}
Z=\sum_i\frac{m_0}{m_i}\Big|\sum_{j=0}^{m_i-1}\tilde\chi_{(i,j)}\Big|^2.
\ee  
Here, for a given $i$ indexing a field $\phi_i$, $\tilde\chi_{(i,j)}$ is a convenient notation for the characters of all the fields along the same orbit of field $\phi_i$ (see Eq. \eqref{eq:orbit}). The integers $m_0$ and $m_i$ are respectively the order of the simple current $\varphi_0$ and the number of distinct primaries along the orbit $i$. Each $m_i$ necessarily divides $m_0$. If $m_i=1$, the orbit contains fixed points of fusing with some simple currents.

For example, consider the WZW model with the chiral algebra $su(2)_8$, which has nine primaries, $0,1,2,\dots,8$. Here, primary $8$ has $h_8=1$ and is the only simple current in the theory. This simple current induces the following integer spin invariant.
\begin{align}
Z_{su(2)_8}&=|\chi_0+\chi_8|^2+|\chi_2+\chi_6|^2+2|\chi_4|^2\\
&=|\tilde\chi_{(0,0)}+\tilde\chi_{(0,1)}|^2+|\tilde\chi_{(1,0)}+\tilde\chi_{(1,1)}|^2 +2|\tilde\chi_{(2,0)}|^2.\nonumber
\end{align}
So in this case, we have $m_1=2$ and $m_2=1$ that implies the primary $4$ is the fixed point of fusing with the simple current $8$.

In the example above, one sees that some original primaries of the CFT under consideration are absent from the corresponding integral spin invariants, namely the fields $1,7,3,5$ in the $su(2)_8$ case. Such absence of fields generally occurs in Eq. \eqref{eq:intSpinModInv}. Now let us make the connection to anyon condensation.

\textbf{\textit{Claim} 1}: The simple currents are the condensates. The fields that are absent from the integer spin invariants of a chiral RCFT are precisely those fields that have nontrivial mutual statistics with the simple currents, and thus confined in the soup of condensates in the framework of Bais. 
It can be explicitly shown that the fields that are not present in the modular invariants possess conformal dimensions that differ from that of the condensates by non-integers. 

It can be shown that these confined fields necessarily decouple from the combination of simple currents under $S$ transformation. The $S$ matrix transforms the chiral characters as
\be\nonumber
\chi_a(-\frac{1}{\tau})=\sum_b S_{ab}\chi_b(\tau),
\ee
where $\tau$ is the conformal parameter. We call the absent fields the \textit{non-local} fields with respect to the simple currents. To show that the non-local fields decouple from the combination of simple currents, namely from $\sum_n^N\chi_{\varphi_0^n}$ amounts to showing that they are absent from $\sum_j^{m_0}\tilde\chi_{(0,j)}(-1/\tau)$ expressed in terms of the $\chi_a(\tau)$'s. 
\begin{align*}
 \sum_{n=1}^N\chi_{\varphi_0^n}(-1/\tau)&=\sum_{n=1}^N\sum_a S_{\varphi_0^n a}\chi_a\\
 &=\sum_a\sum_{n=1}^N S_{\varphi_0^n a}\chi_a\\
 &=\sum_a\sum_{n=1}^N \sum_b \frac{1}{D}\N^b_{\varphi_0^n a}\frac{\theta_b}{\theta_{\varphi_0^n}\theta_a}d_b \chi_a,
\end{align*}
where $\N^c_{ab}$ is the fusion coefficient, i.e., the multiplicity of $c$ in the fusion of $a$ and $b$, $D=\sqrt{\sum_ad_a^2}$ is the total quantum dimension of the chiral RCFT, and $\theta_a=\exp{\ii 2\pi h_a}$ is the self statistics of the primary field $a$. For a simple current $\varphi_0^n$, $\varphi_0^n\x a$ has a unique outcome, say, $b$, and $\N^b_{\varphi_0^n a}\equiv 1$. We thus can conveniently write $\varphi_0^n\x a=a^n$. And since $d_{\varphi_0^n}\equiv 1$, $d_{a^n}=d_a$. Besides, we are dealing with the simple currents $\varphi_0^n$ with integer conformal dimensions. Hence, the above becomes
\begin{align*}
\sum_{n=1}^N\chi_{\varphi_0^n}(-1/\tau)&=\sum_a\sum_{n=1}^N \exp[\ii 2\pi(h_{a^n}-h_a)]\frac{d_a}{D}\chi_a\\
&=\sum_{\phi_i}\sum_{n=1}^N \exp[\ii 2\pi(h_{\phi_{i^n}}-h_{\phi_i})]\frac{d_{\phi_i}}{D}\chi_{\phi_i},
\end{align*}
where we changed the notation to our standard CFT notation of fields, used in particular in Eq. \eqref{eq:orbitCurrentCD}. The second sum  $\sum_{n=1}^N$ in the above becomes the sum over the fields in the orbit $\{\phi_i^n\}$. Using Eq. \eqref{eq:orbitCurrentCD}, we have\begin{align}
\sum_{n=1}^N\chi_{\varphi_0^n}&=\sum_{\phi_i}\sum_{n=1}^N\exp[-\ii 2\pi(\frac{n}{N}\eta\mod \Z)]\frac{ d_{\phi_i}}{D}\chi_{\phi_i}\nonumber\\
&=\sum_{\phi_i}\sum_{n=1}^N\delta_{\eta\!\!\!\! \mod N,0}\frac{ d_{\phi_i}}{D}\chi_{\phi_i},\label{eq:nonlocalAbsence}
\end{align}
where the sum $\sum_{n=1}^N$ is effectively over the representations of $\Z_N$, which naturally leads to the delta function in the second equality. If $\eta\neq 0\mod N$ for some field $\phi_i$, according to Eq. \eqref{eq:orbitCurrentCD}, the fields on the orbit of $\phi_i$ have conformal dimensions different from that of $\phi_i$ by a non-integer. Hence, the above equations manifest that the characters $\chi_{\phi_i}$ of such fields $\phi_i$---in fact of the entire orbit $\{\phi_i^n\}$---are absent from $\sum_{n=1}^N\chi_{\varphi_0^n}$. Note that $\exp[-\ii 2\pi(\frac{\eta n}{N}\mod \Z)]d_{\phi_i}/D$ is precisely the matrix element $S_{\varphi_0^n \phi_{i^n}}$. In addition, for primaries $a$ that appear in the modular invariant, $m_0/m_i$ is exactly the \textit{multiplicity-- } the number of times $a$ splits into a given anyon in the condensed phase. Together this verifies Claim 1. 

The above analysis holds for all unitary RCFTs and the corresponding topological phases as far as simple currents are concerned. This correspondence in turn implies that a child phase has smaller total quantum dimension than its parent phase, which is a special case of formula \eqref{eq:qdrel}.

For topological phases described by the representation categories of $su(2)_k$, simple current condensation can occur only for $k=4l$, $l\in\Z_{\geq 0}$. Such condensation and the corresponding child phases and GDWs are classified by the $D_{2l+2}$ series of Dynkin diagrams and the associated branching matrices. 

As a remark, the result derived above actually does not rely on whether the simple currents have integer conformal dimensions. This is so because in the ratio $\theta_{\phi_{i^n}}/(\theta_{\varphi_0^n}\theta_{\phi_i})$, the conformal dimension in Eq. \eqref{eq:simpleCurrentCD} always cancel out the second term on the RHS of Eq. \eqref{eq:orbitCurrentCD},
we would still obtain Eq. \eqref{eq:nonlocalAbsence}. Nevertheless, in such cases, the simple currents themselves would also decouple from the identity field. As a consequence, the modular invariant so constructed is not an integer spin invariant but merely an automorphism (permutation) invariant.

Another remark\footnote{Thanks to J\"urgen Fuchs for pointing this out to us.} is that in general, there exists modular invariants that do not correspond to Frobenius algebras. Such a modular invariant is thus unphysical, in the sense that it does not describe the torus partition function of a consistent full CFT. Nevertheless, all modular invariants due to simple currents are in fact physical. This is seen by associating a Frobenius algebra with each subset of simple currents that corresponds to a modular invariant.\cite{Fuchs2004a} Ref.\onlinecite{Fuchs2004a} also worked out the representation theory of these algebras. This enabled the proof of various conjectures, e.g., concerning the modular $S$-matrix and boundary conditions, which had been made in Ref.\onlinecite{SCHELLEKENS1990}.

\section{Gapped domain walls and boundaries}\label{sec:GDW}
As pointed out in the end of Sec. \ref{sec:Bais}, the generalized $ADE$ classification encompasses the GDWs between distinct topological phases, GDWs between two copies of the same topological phase, and GBs in one framework. To convey the idea clearly, instead of trying to be most general, let us restrict to the case of $su(2)_k$, which has been thoroughly studied and understood, and whose classifying fusion graphs are truly the Dynkin diagrams of ordinary Lie algebras. We list the $su(2)_k$ $ADE$ classification of modular invariants and their correspondence to anyon condensation in Table \ref{tab:su2_kADE} in the appendix.

In not only $su(2)_k$ but also in other affine Lie algebras, non-simple currents are very rare. In particular in $su(2)_k$, non-simple current condensation only occurs at $k=10$ and $k=28$. For $su(2)_k$, simple currents only occur in cases where $k$ is a multiple of $4$, and all simple current condensation in such cases are classified by the $D_{k/2+2}$ Dynkin diagrams. Note that when $k=16$, there is an exceptional modular invariant of $su(2)_{16}$ characterized by the $E_7$ Dynkin diagram. Nevertheless, this $E_7$ modular invariant does not correspond to any new kind of anyon condensation because it is merely obtained from the $D_{10}$ modular invariant of $su(2)_{16}$ by permutation, namely, by exchanging the $2\oplus 4$ and one copy of the $8$ in the $D_{10}$ invariant. In fact, these two modular invariants differ only by an integer---nine---so, they should give rise to the same RCFT with the chiral algebra in which $su(2)_{16}$ is embedded. This is consistent with the conjecture that according to Bais-Slingerland rules, an anyon condensation in a parent topological phase yields a unique child phase. Besides, mathematically, the Frobenius algebra $A=0\oplus 8\oplus 16$ that leads to this $E_7$ invariant is not commutative nor twist-free\cite{Kirillov2002,Fuchs2002}, which does not directly correspond to anyon condensation in the chiral picture. As such, anyon condensation can serve as a criteria for judging whether a modular invariant is truly exceptional or not. Nevertheless, in the non-chiral picture, anyon condensation corresponding to the $E_7$ invariant can be understood likewise after our discourse on the non-chiral anyon condensation as follows.

Simple current or non-simple current condensation in chiral topological phases always gives rise to a GDW between the parent (larger) topological phase and the child (smaller) topological phase. Note that in chiral cases, a child phase can never be trivial (and we shall get back to this point shortly). This wall can be thought of as a machinery that determines what anyons of the parent phase can enter the child phase and what they will become in the child phase. As brought up in Sec.\ref{sec:Bais}, this machinery can be described by the branching matrix \eqref{eq:bmatrix} of the corresponding vertex operator algebra embedding. The matrix elements of the branching matrix possesses some additional interesting properties beyond what is described previously. Firstly, \be\label{eq:b2mC}
\m_{ab}=\sum_\alpha \b_{\alpha a}\b_{\alpha b},
\ee
which is the very mass matrix for the modular invariant due to the corresponding anyon condensation. Take $su(2)_4$ as a simple example, according to Table \ref{tab:su2_kADE}, the condensation $0\oplus 4$ takes the topological phase to one that corresponds to the RCFT $su(3)_1$. The branching matrix here is
\be
b_{\alpha a}=\bpm
1 & 0 & 0 & 0 & 1\\
0 & 0 & 1 & 0 & 0\\
0 & 0 & 1 & 0 & 0
\epm, 
\ee 
According to Eq. \eqref{eq:b2mC}, we obtain
\be
\m_{ab}=\bpm
 1 & 0 & 0 & 0 & 1 \\
 0 & 0 & 0 & 0 & 0 \\
 0 & 0 & 2 & 0 & 0 \\
 0 & 0 & 0 & 0 & 0 \\
 1 & 0 & 0 & 0 & 1 \\
\epm,
\ee
which yields exactly the $D_4$ modular invariant of $su(2)_4$ in Table \ref{tab:su2_kADE}. Secondly, and more fascinatingly, 
\be\label{eq:b2mU}
\sum_a \b_{\alpha a}\b_{\beta a}=n\mathds{1}+\m_{\alpha\beta},\quad n\in\Z_{\geq 0}.
\ee
Here, $\mathds{1}$ is the identity matrix of the commensurate dimension, and $\m_{\alpha\beta}$ is the mass matrix producing a permutation invariant of the RCFT corresponding to the child phase. Again for $su(2)_4$, we have
\be\label{eq:b2mUsu2_4}
\begin{aligned}
\sum_{a\in su(2)_4} \b_{\alpha a}\b_{\beta a}=
\bpm
 2 & 0 & 0 \\
 0 & 1 & 1 \\
 0 & 1 & 1 \\
\epm
&=\mathds{1}_3+\bpm
 1 & 0 & 0 \\
 0 & 0 & 1 \\
 0 & 1 & 0 \\
\epm\\
&=\mathds{1}_3+\m^{su(3)_1}_{\alpha\beta},
\end{aligned}
\ee
where $n=1$ in this case. This mass matrix $\m^{su(3)_1}_{\alpha\beta}$ gives rise to a permutation invariant of $su(3)_1$, namely $Z_{su(3)_1}=|\chi_0|^2+\chi_3\bar\chi_{\bar 3}+\bar\chi_{\bar 3}\chi_3$, where $0$, $3$, and $\bar 3$ are respectively the scalar, vector and conjugate vector representations of $su(3)$. This has an interesting and important physical meaning. In our previous work\cite{Hung2013,Gu2014a} applying the idea of anyon condensation to construct symmetry-enriched topological (SET) phases, we found that the confined anyons due to the condensation in a parent phase $\C$ can generate a global symmetry group actions on the unconfined phase $\U$ by braiding with the unconfined anyons, and we could work out the explicit representation of such a global symmetry group. Such group actions are usually permutations of certain unconfined anyons in $\U$ and/or symmetry fractionalization of certain unconfined anyons. In our previous work, it was rather involved to solve for the mapping between the Hilbert space basis of $\C$ and that of $\T$ to extract the global symmetry actions. It takes further work to classify such symmetry-enriched phases.

Now, however, we are invited to interpret the matrix $\m^{su(3)_1}_{\alpha\beta}$ in Eq. \eqref{eq:b2mUsu2_4} and in general the $\m_{\alpha\beta}$ in Eq. \eqref{eq:b2mU} as precisely the global symmetry actions on the unconfined phase $\U$ if the confined anyons are indeed able to generate nontrivial permutation actions on $\U$. A case in which the confined anyons cannot generate nontrivial symmetry actions on the unconfined phase is our main example $su(2)_{10}$: using the branching matrix \eqref{eq:su2_10bm} and Eq. \eqref{eq:b2mU}, one can easily find that $\m_{\alpha\beta}^{so(5)_1} =2 \mathds{1}_3$. This is reasonable because physically, it is impossible to mix the fermion $\psi$, whose $d_\psi=1$, with the anyon $\sigma$, whose $d_\sigma=\sqrt{2}$. Since the representation category of $so(5)_1$ is the sibling of the Ising topological phase, there cannot exist nontrivial global symmetry actions on the Ising phase except the global symmetry fractionalization on $\sigma$ and $\psi$ we predicted in Ref.\onlinecite{Hung2013}. A branching matrix however is not able to tell us the possible global symmetry fractionalization of the unconfined anyons. This is because the global symmetry fractionalization is intrinsic to a topological phase\cite{HungWan2013a}. Therefore, a complete classification of symmetry-enriched topological phases looks nigh and deserve our future work.

As mentioned above, in a chiral topological phase, no anyon condensation can completely break the phase to a trivial phase, i.e., $\rep_0 A$ being trivial. This is because $\dim A<\dim\C$ strictly holds for any chiral $\C$ and formula \eqref{eq:qdrel}. In a nonchiral phase as the double $\C\x\overline{\C}$ of a chiral phase $\C$, all the diagonal pairs $a\bar a$ of the elementary anyons $a\in\C$ are self and mutual bosons, and they form a twist-free CSFA $A_{\diag}=\oplus_a a\bar a$. Clearly,
\be
\dim A_\diag=\sum_a d_{a\bar a}=\sum_a d_ad_{\bar a}=\sum_a d_a^2=\dim\C\x\overline\C,
\ee
which, by formula \eqref{eq:qdrel}, implies that $\dim\rep_0 A_\diag=1$, resulting in a trivial unconfined phase. This observation leads us to interpret the diagonal invariants, namely the $A_{k+1}$ series in Table \ref{tab:su2_kADE}, as modular invariants also due to anyon condensation, and such anyon condensation is of the type $A_\diag$ in the doubled topological phases $\C\x\overline\C$. This interpretation is physically equivalent to complete gapping the boundaries of $\C\x\overline\C$ on an open surface.
To fully gap the boundary modes of a doubled phase $\C\x\overline\C$, one has to write down the potential terms of sufficiently many independent boundary fields that can be simultaneously pinned to certain vacuum expectation values. The set of such boundary fields comprise what is known as a Lagrangian algebra\cite{Levin2013,Kitaev2012,Kong2013,HungWan2014}. Other independent boundary fields absent from this subalgebra are nonlocal with respect to at least one element of the subalgebra and thus will be confined if the Lagrangian algebra is condensed. A twist-free CSFA $A_\diag$ is precisely a Lagrangian algebra. A completely gapped boundary is dubbed a GB.

As explained previously, anyon condensation induces a modular invariant of the RCFT corresponding to the unconfined phase surviving the condensation. And here in the case of doubled phases, an anyon condensation $A_\diag$ results in a trivial phase, so the induced modular invariant should contain only terms corresponding to the condensed anyons, which is precisely one in the form of the $A_{k+1}$ series in Table \ref{tab:su2_kADE}. This is consistent with the modular invariants due to anyon condensation in chiral topological phases. As such, our interpretation bears no ambiguity or confusion. The discussion above can be generalized to modular tensor categories as representation categories of other affine Lie algebras or vertex operator algebras than just $su(2)_k$. Therefore, it is justified to claim that the diagonal modular invariants of the chiral RCFTs corresponding to a chiral phases $\C$ classify certain GBs of the doubled topological phases $\C\x\overline\C$.

What remains now in Table \ref{tab:su2_kADE} is the $D_{2l+1}$ series of permutation invariants, which are not diagonal but do not appear to be induced by any anyon condensation in the corresponding chiral topological phases. Nevertheless, by an argument similar to that about the diagonal invariants, the $D_{2l+1}$ series can also be understood as due to anyon condensation in the doubled phase $su(2)_k\x\overline{su(2)_k}$, for $k=4l-2$, and such anyon condensation is characterized by the twist-free CSFA
\be\label{eq:Aperm}
\begin{aligned}
A_{\mathrm{perm}}=&\bigoplus\limits_{n=0}^{2l-1} (2n, \,\, \overline{2n})\oplus (2l-1,\,\, \overline{2l-1})\\
 &\bigoplus\limits_{n=1}^{2l-3}(n,\,\,\,\overline{4l-2-n})\oplus (4l-2-n,\,\,\bar n)
\end{aligned},
\ee
where each pair $(m,\bar{n})$ denotes an anyon corresponding to the direct product of the chiral sector $m$  and anti-chiral sector $\bar{n}$. In fact, the permutation condensation $A_\mathrm{perm}$ is also a Lagrangian algebra that can completely destroy the doubled phase or gap all the boundary modes. Here is the reason. First, a pair of anyon or fields that can be permuted must have the same quantum dimension and topological spins differ by merely an integer, such as the pair of $n$ and $4l-2-n$ in $su(2)_k$, as in Eq. \eqref{eq:Aperm}. Second, a permutation invariant or the corresponding $A_\mathrm{perm}$ contain all the elementary fields/anyons. Hence, one sees from Eq. \eqref{eq:Aperm} and even more generally,
\be
\dim A_\mathrm{perm}=\dim \C\x\overline\C,
\ee
indicating a trivial $\rep_0 A_\mathrm{perm}$ or unconfined phase.

Therefore, $A_\diag$ and $A_\mathrm{perm}$ both correspond to the GBs of doubled topological phases $\C\x\overline\C$ on open surfaces. In particular, for $\C$ representation categories of affine Lie algebras $\mathfrak{g}_k$, or direct products of these algebras, the GBs are fully classified by $A_\diag$ and $A_\mathrm{perm}$, or the corresponding modular invariants. The branching matrices in such cases are obtained as in the chiral case.

\section{Quantum Double of a Gauge Theory}
In the special case where the topological phase is describable by a quantum double with group $G$, which
are examples of lattice gauge theories, there are well known \emph{Lagrangian algebra}, corresponding to condensation of all electric charges or all magnetic charges, that are guaranteed to lead to a gapped boundary separating the phase from the trivial vacuum. 

In those cases, it is possible to derive commutativity relation from our knowledge of the $S$ matrix in these models and well known relations satisfied by representations of groups.

To begin with, let us recall the form of the $S$ matrix for the quantum double of a finite gauge group $G$,
\be
S_{(A,\mu) \, (B,\nu)} = \frac{1}{|G|}\sum_{g\in C^A,\, h\in C^B,\, [g,h] = 1} \overline{\chi^{Z^A}(h)_{\mu} \chi^{Z^B}(g)_{\nu}},
\ee
where $C^A,C^B$ denote the conjugacy classes of $G$, $Z^A, Z^B$ the corresponding centralizers, and $\mu,\nu$
the representations of $Z^A,Z^B$ respectively. 

Consider then the ``electric'' Lagrangian algebra corresponding to the condensing all the pure electric charges, which are guaranteed to behave like self-bosons. 

In this case, Bais-Slingerland rules map the condensed anyons all to the vacuum, and what are not condensed are confined. This map can be described by a matrix analogous to the branching matrix \eqref{eq:bmatrix}. Let us again take this $\b$ matrix notation to emphasize the generality and universality of anyon condensation. Note however that in the current case, the $\b$ matrix is simply a single-row matrix, and let us write it as  $\b_{1(A,\mu)}$. According to Bais-Slingerland rules, the nonvanishing components of $\b$ are $\b_{1(e,\mu)}=d_\mu$, the dimension of the representation $\mu$ of $G$, where $(e,\mu)$ denotes a pure electric charge, with $e\in G$ the identity element. All dyons and magnetic charges carrying non-trivial labels of conjugacy classes are confined and do not appear in $\b$.

Now let us check for commutativity:

\begin{eqnarray}
&&\sum_\mu \b_{1(e,\mu)} S_{(e, \mu) (B,\nu)} = \frac{1}{|G|}\sum_\mu d_\mu \sum_{h\in C^B} \overline{\chi^{G}_{\mu}(h) \chi^{Z^B}_\nu(e)} \nonumber \\
&&= \frac{d_\nu |C^B|}{|G|} \sum_\mu d_\mu \overline{\chi^{|G|}_\mu(h)} = \frac{d_\nu |C^B|}{|G|} \delta_{e h} |G| \nonumber \\ 
&&= \b_{1(B,\nu),}
\end{eqnarray}
where we have made use of properties of characters of any finite group in the 
second equality. Amazingly, the commutativity condition is well aware of group representation theory.

Alternatively, another set of well known Lagrangian algebra is the ``magnetic'' condensate, in which all the pure magnetic charges characterized by a conjugacy class and the trivial representation condenses all at once. 

In which case, the non-zero components of the $\b$ matrix is given by $\b_{1 (A,1)}= 1$ (i.e. only one linear combination of the $|C^A|$ members of the conjugacy class can take part in the condensation.). The commutativity condition is then reduced to
\begin{eqnarray}
&&\sum_A \b_{1(A,1)} S_{(A, 1) (B,\nu)} \nonumber \\
&&= \frac{1}{|G|}\sum_A   \sum_{h\in C^B, g\in Z^B  \& g\in C^A} \chi^{Z^B}_\nu(g) \nonumber \\
&&= \frac{1}{|G|} \sum_{h\in C^B} \sum_{g\in Z^B} \chi^{Z^B}_\nu(g) \nonumber \\
&&=\frac{1}{|G|} |C^B| |Z^B| \delta_{\nu 1}  \nonumber \\
&&= \delta_{\nu 1} \equiv \b_{1,(B,1)}
\end{eqnarray}
Again the results following from Bais-Slingerland rules again implies commutativity of the $\b$ matrix with the $S$ matrix.

\section{Conclusions}
For a possible systematic classification of the anyon condensation and thus GBs and GDWs, it is crucial to have a rigorous mathematical definition of these physical phenomena, and these have been proposed notably in Ref.\onlinecite{Kitaev2012,Fuchs2014,Kong2013}. The construction however is very different from those in Ref.\onlinecite{Bais2002,Bais2009,Bais2009a}, which is motivated by purely physical consideration. In this paper, we fill the gap by recovering every important ingredient in Ref.\onlinecite{Bais2002,Bais2009,Bais2009a} making use of the properties of twist-free CSFA in a UMTC, strengthenging the connection between the mathematics and the physical intuition, and thus allowing for the existing classification, namely a generalized $ADE$ classification, of the twist-free CSFA to be a direct classification of possible GBs and GDWs in many known phases. We also explore how far one can recover the data of a twist-free CSFA, notably a modular invariant, starting from Bais-Slingerland rules. We find that at least for simple currents condensation and electric/magnetic condensates in any quantum doubles of a group $G$ they are in 1-1 correspondence. We also connect these results with the novel proposal of \cite{Lan2014} that gives a very simple set of rules to classify the same objects. It is an on-going programme, both of physical and mathematical interest to find the minimal set of data to identify GBs and GDWs. 

Anyon condensation appears to be related to gauging the generalized global symmetry recently proposed by Gaiotto \textit{et al}.\cite{Gaiotto2014} We shall report our investigation in this respect elsewhere.

\begin{acknowledgements}
We are indebted to J\"urgen Fuchs and Liang Kong for invaluable comments on the manuscript. We thank Davide Gaiotto, Yuting Hu, Liang Kong, Tian Lan, and Xiao-Gang Wen for helpful discussions.  Y.W. appreciates Davide Gaiotto particularly for having inspired him to look deeper into the connection between anyon condensation and vertex operator algebra embedding. Y.W. is supported by the John Templeton Foundation. Research at Perimeter Institute is supported by the Government of Canada through Industry Canada and by the Province of Ontario through the Ministry of Economic Development \& Innovation.
\end{acknowledgements}

\begin{appendix}
\section{Bais-Slingerland Rules}
Here we recapitulate Bais-Slingerland rules of anyon condensation as how we understand them.
\begin{enumerate}
\item Anyon condensation of a topological phase $\C$ selects a set of self and mutually local anyons $\{\gamma\}$ as a subset of all the elementary anyons (i.e., simple objects) of $\C$. Note that this set always includes the trivial anyon or vacuum $1$.
\item If an anyon $a$ of $\C$ is a (meta) fixed point of $m$ condensed anyon $\gamma_i$, $i=1,\dots,m$, namely $a$ appears again in the fusion product $a\x \gamma_i$, $\forall i$, $a$ will split into $m$ species (not necessarily all different) in the condensed phase:
\be\label{eq:split}
a\rightarrow\sum_{i=1}^m n^i_a a_i,
\ee
where $n_a^i\in\Z_{\geq 0}$ is the multiplicity of species $a_i$. Clearly, $\sum_{i=1}^m n^i_a=m$. Note that a condensate $\gamma$ may also split.
\item The above splitting preserves quantum dimension:
\be\label{eq:splitQDconserve}
d_a=\sum_{i=1}^m d_{a_i}.
\ee
\item Splitting and fusion commute:
\be\label{eq:splitFusionCommute}
\biggr(\sum_i n^i_a a_i\biggr)\x \biggr(\sum_j n^j_b b_j\biggr) =
\sum_{c,k}N^c_{ab} n^k_c c_k.
\ee
\item If two anyons $a$ and $b$ are related by fusing with a condensate $\gamma$, i.e., $a=b\x\gamma$,  they should be identified as a single species in the condensed phase. Note that more than two anyons can be identified, and such identification may be restricted to the split components.
\item If the anyons being identified have the topological spins different by merely integers, the anyon species as the identification of them would inherit their topological spin modulo the integers and is an unconfined anyon in the condensed phase. Otherwise, the identification leads to a confined anyon in the condensed phase. 
\item The condensed phase including both confined and unconfined anyons is called the $\T$ phase, whereas that consists of unconfined anyons only is called the $\U$ phase. It is conjectured that if the original phase is a UMTC, $\U$ is also a UMTC.
\end{enumerate}
The above rules can be easily and systematically applied for simple currents condensation; however, for nonsimple currents condensation, their application is rather invovled, in particular the rule 2 of splitting and rule 5 of identification are mingled together, which requires careful case by case study.

\section{Twist-free commutative separable Frobenius algebra}
The concept of twist-free commutative separable Frobenius algebra (CSFA) of a UMTC $\C$ has been employed to classify the quantum subgroups of quantum groups, whose representation categories are UMTCs, as well as the embeddings of vertex operator algebras, whose representation categories are also (not necessarily unitary) MTCs.\cite{Kirillov2002} Despite the complexity of the mathematics involved, we shall give a brief account of this concept in physical terms. We refer the reader to Ref.\onlinecite{Kirillov2002,Fuchs2002,Kong2013} for more systematic discussions on twist-free CSFAs. 

\begin{itemize}
\item {\bf Frobenius algebra.}\,\,\,Note that a Frobenius algebra is not only an algebra but also a coalgebra; however, we shall not need its coalgebra aspect for our purposes. Hence, in this appendix and throughout the paper, we treat a Frobenius algebra as only an associative algebra. A Frobenius algebra $A$ is an object in $\C$. In general $A$ is not a simple object but a direct sum of simple objects. For a topological phase described by $\C$, the simple objects are the distinct elementary anyon types. This object $A$ is an algebra because it is equipped with a product $\mu:A\ox A\rightarrow A$ and an inclusion $\iota_A:1_\C\hookrightarrow A$, where $1_\C$ is the unit object or vacuum of $\C$, such that $1_\C$ is also the unit of $A$. The unit is also required to be unique, namely $\dim \Hom_\C(1_\C,A)=1$. Such uniqueness is called \textit{haploid} in mathematics. The product $\mu$ is {\bf\textit{associative}} and {\bf\textit{commute}} with the braiding of $A$. The former is the associativity. The latter means $\mu\circ R_{AA}=\mu$, where $R_{AA}$ is the $R$ matrix of $A$ in $\C$. This commutativity is physically sound because $A$ is going to become the new vacuum when it condenses. Let us formally write $A=1\oplus\Upsilon$, where $\Upsilon$  collectively denotes the direct sum of other simple objects of $\C$ that may appear in $A$. The algebra object $A$ is self-dual, also physically consistent with that $A$ is going to be the new vacuum. This is also called {\bf \textit{rigidity}} of $A$ in Kirillov\cite{Kirillov2002}. 

\item{\bf representation category Rep$A$}. A CSFA $A$ induces a representation category $\rep A$ over $A$. In order that the representations over $A$ are semisimple, $A$ is required to be {\bf \textit{separable}}. This allows a well-defined tnotion of simple objects in $\rep A$ and the nonsimple objects as direct sums of the simple ones. This is a key notion for anyons in topological phases to be well-defined. The category $\rep A$ is a tensor category if $A$ is {\bf \textit{twist-free}}, i.e., $\theta_A=\id_A$, where $\theta_A$ can be understood as the self-statistics of $A$ as an object in $\C$. A trivial example of a twist-free CSFA is $A=1_\C$ in any $\C$, which is called \textit{transparent}.

The category $\rep A$ in general is not a braided tensory category; however, it has a subcategory that is braided. This subcategory is denoted by $\rep_0 A$ and consists of the objects in $\rep A$ whose fusion with $A$ commutes with its braiding with $A$. Formally we can write this as
\be
\rep_0 A=\{X\in\T|(A\otimes_A X)R_{XA}R_{AX}=A\otimes_A X\},
\ee 
where the fusion $\ox_A$ is defined with respect to $A$. We leave the detail of this definition to the references. This is shown in Ref.\cite{Kirillov2002,Frohlich2006}.

We note that the definitions of a twist-free CSFA in Ref.\onlinecite{Kirillov2002} and Ref.\onlinecite{Fuchs2002} are not identical but as discussed in Ref.\onlinecite{Kong2013}, they are equivalent.

\end{itemize}
\section{Modular data and $ADE$ classification of $su(2)_k$}
We first tabulate in Table \ref{tab:su2_10} some of the topological data of $su(2)_{10}$ we extensively used in the main text.
\begin{table}
\begin{tabular}{|c|c|c|}\hline
sectors $a$ & $d_a$ & $h_a$ \\\hline
$0$ & $1$ & $0$ \\[0.5ex]
$1$ & $\sqrt{2+\sqrt{3}}$ & $1/16$ \\[0.5ex]
$2$ & $1+\sqrt{3}$ & $1/6$ \\[0.5ex]
$3$ & $\sqrt{2}+\sqrt{2+\sqrt{3}}$ & $5/16$ \\[0.5ex]
$4$ & $2+\sqrt{3}$ & $1/2$ \\[0.5ex]
$5$ & $2\sqrt{2+\sqrt{3}}$ & $35/48$ \\[0.5ex]
$6$ & $2+\sqrt{3}$ & $1$ \\[0.5ex]
$7$ & $\sqrt{2}+\sqrt{2+\sqrt{3}}$ & $21/16$ \\[0.5ex]
$8$ & $1+\sqrt{3}$ & $5/3$ \\[0.5ex]
$9$ & $\sqrt{2+\sqrt{3}}$ & $33/16$ \\[0.5ex]
$10$ & $1$ & $5/2$ \\\hline
\end{tabular}
\caption{$su(2)_{10}$ topological sectors, quantum dimensions, and topological spins.}\label{tab:su2_10}
\end{table}

Table \ref{tab:su2_10} and that of more general chiral algebras $su(2)_k$ can be obtained using several formulae. Although we only considered $k=10$ and $k=8$ in our examples, for completeness we consider general $k\in\Z$ in this appendix. These formulae can be found in other references too, e.g., Ref.\onlinecite{Eliens2013,Fuchs2002}.

An affine Lie algebra $su(2)_k$ has $k$ distinct topological sectors, $0,1,\dots,k$, where $0$ is the trivial or vacuum sector. The fusion algebra of these sectors reads
\be\label{eq:su2kfusion}
a\x b=c_{ab}+(c_{ab}+2)+\cdots+\min\{a+b,2k-a-b\},
\ee
where $c_{ab}=|a-b|$. The multiplicity $N^c_{ab}=1$ if $|a-b|\leq c\leq \min \{a+b,2k-a-b\}$, $a+b+c=0\pmod{2}$, and $a+b+c\leq 2k$; otherwise, $N^c_{ab}=0$. A sector $a$ has quantum dimension and topological spin respectively
\be\label{eq:su2kqdTheta}
\begin{aligned}
d_a=\frac{\sin(\frac{a+1}{k+2}\pi)}{\sin\frac{\pi}{k+2}},\\
h_a=\frac{a(a+2)}{4(k+2)}.
\end{aligned}
\ee
The $R$ matrix elements read
\be\label{eq:su2kR}
R^{ab}_c=\ii^{c-a-b}q^{\frac{1}{8}[c(c+2)-a(a+2)-b(b+2)]}.
\ee
The $F$ symbols are matrices:
\be\label{eq:su2kF}
\Fm{a}{b}{c}{d}{e}{f}=\ii^{a+b+c+d}\sqrt{\frac{d_ed_f}{d_ad_d}[a+1]_q[d+1]_q} \sixj{c}{e}{a}{b}{f}{d}^*,
\ee
where the $6j$ symbols
\begin{align*}
\sixj{c}{e}{a}{b}{f}{d}&=\Delta(c,e,a)\Delta(e,c,d)\Delta(e,b,d)\Delta(c,d,f)\\
\x \sum_z\Biggl\{ & \frac{(-1)^z[z+1]_q!}{[z-\frac{c+e+a}{2}]_q![z-\frac{a+b+f}{2}]_q! [z-\frac{e+b+d}{2}]_q!} \\
  &\x\frac{1}{[z-\frac{c+d+f}{2}]_q![\frac{c+e+b+f}{2}-z]_q!}\\
  &\x\frac{1}{[\frac{c+a+b+d}{2}-z]_q![\frac{e+a+f+d}{2}-z]_q!}\Biggr\}
\end{align*}
are defined with
\[
\Delta(a,b,c)=\sqrt\frac{[\frac{-a+b+c}{2}]_q![\frac{a-b+c}{2}]_q![\frac{a+b-c}{2}]_q!} {[\frac{a+b+c}{2}+1]_q!},
\]
which is invariant under permutation of its variables, and the $q$-numbers and $q$-factorials
\[
[n]_q=\frac{q^{n/2}-q^{-n/2}}{q^{1/2}-q{-1/2}},\quad [n]_q!=\prod_{m=1}^n[m]_q.
\]
By definition $[0]_q!\equiv 1$. Note that the sum over $z$ appeared in the expression of the $6j$ symbols is carried from $\max\{\frac{c+e+a}{2},\frac{a+b+f}{2},\frac{e+b+d}{2}, \frac{c+d+f}{2}\}$ to $\min\{\frac{c+e+b+f}{2},\frac{c+a+b+d}{2},\frac{e+a+f+d}{2}\}$.

Now we tabulate the $ADE$ classification of $su(2)_k$ modular invariants as follows.
\begin{widetext}

\begin{table}[ht]
\begin{tabular}{|c|c|c|c|c|}\hline
& Anyon condensation & Series & Level $k$ & Modular Invariants \\\hline
GBs& \pbox{10cm}{No condensation in the chiral case\\or diagonal condensation $A=\oplus_{n=0}^kn\bar n$ \\ in the nonchiral pair $su(2)_k\x \overline{su(2)_k}$ }& $A_{k+1}$ & arbitrary & $\sum\limits_{n=0}^k |\chi_n|^2$\\\hline
GDWs& \pbox{10cm}{Simple current condensation\\ $A=0\oplus k$} & $D_{2l+2}$ & $4l$ & $\sum\limits_{n=0}^{2l-2}|\chi_n+\chi_{4l-n}|^2+2|\chi_{2l}|^2$ \\\hline
GBs& \pbox{10cm}{No condensation in the chiral case\\or\ condensation $A=\bigoplus\limits_{n=0}^{2l-1} (2n, \overline{2n})\oplus(2l-1, \overline{2l-1})$\\ $\bigoplus\limits_{n=1}^{2l-3}(n,\,\,\overline{4l-2-n})\oplus (4l-2-n,\,\,\bar n)$\\ in the nonchiral pair $su(2)_k\x \overline{su(2)_k}$ } & $D_{2l+1}$ & $4l-2$ & \pbox{10cm}{$\sum\limits_{n=0}^{2l-1}|\chi_{2n}|^2+|\chi_{2l-1}|^2$ \\$+\sum\limits_{n=1}^{2l-3}(\chi_n\bar\chi_{4l-2-n}+\chi_{4l-2-n}\bar\chi_n)$} \\\hline
GDWs& \pbox{10cm}{Non-simple current condensation\\$A=0\oplus 6$} & $E_6$ & $10$ & $|\chi_0+\chi_6|^2+|\chi_3+\chi_7|^2+|\chi_4+\chi_{10}|^2$ \\\hline
GDWs& Permutation of the $D_{10}$ invariant of $su(2)_{16}$ & $E_7$ & $16$ & \pbox{10cm}{$|\chi_0+\chi_{16}|^2 +|\chi_4+\chi_{12}|^2+|\chi_6+\chi_{10}|^2+|\chi_8|^2$\\$+\chi_8(\bar\chi_2+\bar\chi_{14}) +(\chi_2+\chi_{14})\bar\chi_8$} \\\hline
GDWs& \pbox{10cm}{Non-simple + simple current condensation\\$A=0\oplus 10\oplus 18\oplus 28$} & $E_8$ & $28$ & $|\chi_0+\chi_{10}+\chi_{18}+\chi_{28}|^2+|\chi_6+\chi_{12}+\chi_{16}+\chi_{22}|^2$ \\\hline
\end{tabular}
\caption{The $ADE$ classification of $su(2)_k$ modular invariants and their correspondence with anyon condensation. The branching matrix in each case can be easily read off from the table. In the table, we assume $n\in \Z$.}
\label{tab:su2_kADE}
\end{table}
\end{widetext}

\section{Some collected observations involving simple currents condensation}
Consider for simplicity chiral topological phases $\C$. For simple current condensation, Bais-Slingerland rules can easily lead to precise consequences. Since a simple current is a self-boson, if it condenses, then all the simple currents generated by fusion with the condensed simple current in the same phase will condense too. Hence, all the condensates form a closed set under fusion; if fusion is taken as a group multiplication, this closed set can be identified with a cyclic group $\Z_{|c_0|}$, where as defined above $|c_0|$ is the order of the unit simple current $c_0$ of the phase. In more general cases where a topological phase is the direct product of, say, $m$  topological phases, the simple currents form the Abelian group $\Z_{N_1}\x\Z_{N_2}\x\cdots\x\Z_{N_m}$.   If $a$ is a fixed point of fusing with some $c_0^p$, with $1\leq p\leq |c_0|-1$ an integer, then there are seemingly two cases. First, if $|c_0|/p=q\in\Z$, then $a$ is also a fixed point of fusing with the simple currents $c_0^p,c_0^{2p},\dots,c_0^{(q-1)p}$. Second, if $p>1$ and $|c_0|/p\notin \Z$, then $a$ is a fixed point of all the simple currents in the phase; hence, this case is equivalent to the case where $a$ is a fixed point of $c_0$. Therefore, both cases can be merged into one  where $|c_0|/p\in\Z$ is assumed, which obviously includes the possibility of $p=1$. 

By unitary,
i.e., a topological sector and its anti-sector have a unique way of fusing into the vacuum, it can be easily shown that as long as $a$ is a fixed point of a simple current condensate $c_0^p$, it must split as
\be 
a\rightarrow \sum_i n^a_i a_i,
\ee
where $a_i$'s are the sectors (possibly confined) in the broken phase, and $n^a_i$ is the multiplicity of $a_i$ in this splitting. The quantum dimensions are conserved as
\be
d_a=\sum n^a_i d_{a_i}.
\ee
And $h_{a_i}=h_a$ for all $i$. 

On the other hand, two sectors (including the sectors obtained by splitting some sectors of $\A$), $a$ and $b$ should be identified as a single sector in the broken phase if $a\x c_0^{m p}=b$ for any integer $1\leq m\leq q-1$ because they become indistinguishable in the presence of the new vacuum. An interesting consequence is, if $h_b-h_a\notin \Z$, the resultant sector that identifies $a$ and $b$ ought to be confined in the broken phase. This is consistent to the fact that if a sector has non-trivial monodromy with a condensate, it has to pull a string, whose energy is proportional to its length, when it moves around in the new vacuum and thus is confined. To be precise, we let $\gamma=c_0^{mp}$ for some integer $m$ such that $a\x\gamma=b$. The monodromy between $a$ and $\gamma$ reads
\be
M_{a\gamma}=\frac{S_{a\gamma}S_{11}}{S_{a1}S_{\gamma 1}}=\sum_b N^b_{a\gamma} \frac{\theta_b d_b}{\theta_a\theta_\gamma d_a}=\frac{\theta_b}{\theta_a}=\e^{\ii 2\pi(h_b-h_a)}.
\ee
Here, use is made of $S_{a1}=d_a/D$, $S_{11}=S_{\gamma 1}=1/D$, $S_{a\gamma}=\sum_b N^b_{a\gamma}\theta_bd_b/(\theta_a\theta_\gamma D)$, $d_\gamma=1$, and $a\x\gamma=b$ that implies that $N^b_{a\gamma}=1$ and $d_a=d_b$.
Clearly, $M_{a\gamma}$ is non-trivial if and only if $h_b-h_a\notin\Z$.
The above equation immediately implies that if $a\x\gamma=a$, $M_{a\gamma}=1$. That is, a fixed point of a simple current condensate, despite splitting, yield only unconfined sectors in the broken phase.

One can see that two unconfined sectors can never fuse to a confined sector. This is because physically, a distant observer is not able to distinguish the system containing two sectors from the one containing the fusion product of the two sectors; were the two sectors unconfined, i.e., were they mutually local with respect to the vacuum, there would be no way for their fusion product to have nontrivial monodromy with the vacuum. Therefore, the unconfined sectors in the broken phase are closed under fusion and comprise a well-defined topological phase on their own. We call this unconfined phase $\U$.
As quantum groups, we have $\U\subset\C$.

Consider the follwing example. The topological phase $\C$ is characterized by the quantum group $\U_q(su(2))$ with $q=\exp(\ii\pi/3)$. The corresponding CFT has the chiral algebra $su(2)_4$. This phase $\C$ has five topological sectors, $0,1,2,3$, and $4$. The only simple current in this spectrum is sector $4$. The fusion with sector $4$ has a single fixed point, which is sector $2$, as $2\x 4=2$. Hence, if sector $4$ condenses, sector $2$ would have to split. Since $d_2=2$, it has to split into two pieces, $2_1$ and $2_2$, which inherits the topological spin of sector $2$, which is $h_2=1/3$.

\section{Some thoughts on $W$ matrices}
In a recent work, Lan \textit{et al}\cite{Lan2014} proposes the idea of using what they call a $W$ matrix, whose entries are natural numbers, to characterize the GDWs between any two topological phases and the GBs of any topological phases. They offer a set of consistency conditions of a $W$ matrix. If for two topological phases, the set of conditions bear no solution of $W$, there does not exist a GDW between these two phases. If a topological phase cannot have a GB, there is no solution of $W$ either. We leave the details of these consistency conditions to the cited paper. As mentioned in the main body of the paper, there is a relation between anyon condensation and $W$ matrices. For example, a branching matrix and a mass matrix may be regarded as $W$ matrices too. In all the examples we have studied, we have found a one-to-one correspondence between an anyon condensation and a $W$ matrix. At present it is still unclear whether these $W$ matrices are in 1-1 correspondence with a CSFA. In this appendix, however, we report some of our findings of the $W$ matrices. We would like to dwell on the implications of commutativity between a $W$ matrix and the modular $S$ and $T$ matrices of the two topological phases on the two sides of the GDW. This is the same commutativity as that in Eq. \eqref{eq:STbST}, with the $\b$ matrix which is in fact a $W$ matrix.

\subsection{Gapped domain wall and global symmetry}
The commutativity between the $W$ matrix that characterizes a gapped domain wall between two phases and the $S$ matrices of the two phases implies that the $W$ matrix may be intimately related to global symmetries. In particular, when the two phases separated by a GDW are the same phase, the wall, associated with which the $W$ matrix is a square matrix, can be thought of as implementing a global symmetry on the phase. Here the commutativity reads 
\be \label{eq:commutS=B}
WS_B=S_BW.
\ee 
This is plausible because the $W$ matrix serves as an endomorphism on the set of topological sectors. This is precisely what a global symmetry does on a topological phase, and the symmetry should also commute with the physical observables of the phase. 

On the other hand, a parent phase $A$ can not only produce a child phase $B$ by its anyon condensation but also generate a global symmetry on phase $B$ by the braiding between $B$'s sectors and the confined sectors of $A$ due to the condensation. Since the generated global symmetry is explicitly represented on $B$'s sectors, from such a symmetry representation, we can obtain a gapped domain wall $W$ between two copies of $B$. 

Let us consider the example where phase $A$ is the \II and $B$ the $\Z_2$ toric code. We know that by condensing the $\psi\bar\psi$ sector, $A$ breaks into $B$. The generated global symmetry is $\Z_2$ represented by the $2\x 2$ identity matrix $I_2$ and the Pauli $\sigma_x$ matrix\cite{Gu2014a}. The identity $I_2$ acts on the trivial sector $1$ and the fermion $\epsilon$ of the $\Z_2$ toric code, whereas $\sigma_x$ acts on the subspace as the direct sum $e\oplus m$. If we order $B$'s sectors as $1,e,m,\epsilon$, we can reorganize the $I_2$ and $\sigma_x$ into a $4\x 4$ matrix, which represents an endomorphism on $B$, namely
\[
W=\bpm
1 & 0 & 0 & 0\\
0 & 0 & 1 & 0\\
0 & 1 & 0 & 0\\
0 & 0 & 0 & 1
\epm.
\] 
This is exactly the nontrivial gapped domain wall between two copies of $\Z_2$ toric code found in Ref[\onlinecite{Lan2014}]. One can easily check the commutativity between this $W$ and the $S$ matrix of $\Z_2$ toric code.

As another example, we know the chiral topological phase $SU(2)_8$, via condensing its sector $8$, can break into the $\text{Fibo}\x\text{Fibo}$ phase and generate a $\Z_2$ symmetry on the latter. It turns out that this $\Z_2$ symmetry is also represented by the $W$ matrix above. Moreover, this $W$ matrix indeed commutes with the $S$ matrix of the $\text{Fibo}\x\text{Fibo}$ phase, we can thus infer that this $W$ characterizes a gapped domain wall between two copies of the $\text{Fibo}\x\text{Fibo}$ phase. As such, we need not to solve the commutativity equation for $W$ but simply obtain the result using anyon condensation.

The discussion above again hints that a $W$ matrix between two phases $A$ and $B$ encodes in an intricate way the information of the confined sectors due to the breaking of $A$ into $B$, although apparently $W$ has only zero entries corresponding to the confined sectors.

Furthermore, write the commutation \eqref{eq:commutS=B} explicitly as
\be
\sum_bW_{ab}S_{bc}=\sum_dS_{ad}W_{dc},
\ee
and let $a=1$ in herein, we get 
\be
1=S_{1c}=\sum_b\delta_{1b}S_{bc}=\sum_dS_{1d}W_{dc}=\sum_d W_{dc},\ \forall c.
\ee
Here, use is made of $W_{1b}=\delta_{1b}$, which is true because there is no anyon condensation in this case. Since $W$'s entries are all non-negative integers, we can conclude that in this case each row and and each column of any $W$ in this case contains one and only one nonzero value, which is unity. This is consistent to the fact that a gapped domain wall between the same phases must not mediate any splitting or identification of the topological sectors and that the wall implements a global symmetry. This appears to put a strong constraint on how a global symmetry group on a topological phase may act on, or in other words, may be represented on, the spectrum of the phase. One may find the resemblance between such seemingly existing constraint on global symmetries and our discussion below Eq. \eqref{eq:b2mU}.   
\subsection{Modular invariance} 
Commutativity of $W$ with the modular matrices implies modular invariance. The $S^{A,B}$ corresponds to $S$ matrices of the two phases connected by a gapless domain wall defined by $W$.
Consider having 
\be
Z(\tau) = \sum_{a,i} W_{ai} Z_a \bar{Z}_i
\ee
Under modular transformation
\begin{eqnarray}
Z(-1/\tau)&& = \sum_{a,b,i,j} W_{i a} S_{ab} Z_{b} \bar{Z}_k S^{\dag}_{k i}\nonumber \\
&&=  \sum_{b,i,k} S_{ik}W_{k b} Z_{b} \bar{Z}_k S^{\dag}_{k i} \nonumber \\
&&=  \sum_{b,k} W_{k b} Z_{b} \bar{Z}_k = Z(\tau),
\end{eqnarray}
where we have made use of 
\be
\sum_a W_{i a} S_{ab} = \sum_{k} S_{ik}W_{k b},
\ee
and that we have assumed the unitarity of $S_{ab}$ and $S_{ij}$.

For invariance under $T$ transformation, we have
\be
Z(\tau+1) = \sum_{a,b} W_{i a} \exp(2\pi i (h_a- h_i)) Z_a \bar{Z}_i,
\ee
which implies invariance for 
\be
\theta_a= \theta_i,
\ee
which is indeed the case.

\subsection{Some equalities}
Using the explicit expression for the $S$ matrix
\be\label{eq:Smatrix1}
S_{ab} = \frac{1}{D} \sum_c N^c_{ab} \frac{\theta_c d_c}{\theta_a \theta_b},
\ee
commutativity also implies
\be
\frac{1}{D_A}\sum_{c, a}W_{i a} N^{c}_{ab} \frac{\theta_c d_c}{\theta_a \theta_b}
= \frac{1}{D_B}\sum_{j,k} N^{k}_{ij} W_{j b}\frac{\theta_k d_k}{\theta_i \theta_j}
\ee
Now, consider the special case $b=1$,
then we have
\be
\frac{1}{D_A}\sum_{c, a}W_{i a} \delta_{a,c} \frac{\theta_c d_c}{\theta_a }
=  \frac{1}{D_B}\sum_{j,k} N^{k}_{ij} W_{j 1}\frac{\theta_k d_k}{\theta_i},
\ee 
where we have made use of the fact that $\theta_b= \theta_1=1= \theta_j$ for $W_{j1}$ non vanishing. 
Then we end up with
\be
 \frac{1}{D_B}\sum_{j,k} N^{k}_{ij} W_{j 1}\frac{\theta_k d_k}{\theta_i}= \frac{1}{D_A}\sum_{a}W_{i a}  d_a.
\ee
Choose also $i = 1$. We have then
\be
 \frac{1}{D_B}\sum_{j}W_{j 1}d_j = \frac{1}{D_A}\sum_{a}W_{1 a}  d_a.
\ee

Now we make the input that phase $B$ is condensing to phase $A$. We implicitly assume that $D_B> D_A$, and that $W_{1a}= \delta_{1a} $. Then we are left with
\be\label{eq:DB/DA}
\frac{D_B}{D_A} = \sum_{j}W_{j 1}d_j .
\ee
This can be easily checked to be the case in all our cases of anyon condensation.

Moreover, this can be compared with and equation obtained in Ref.\onlinecite{Bais2009a},
\be
\frac{D_B}{D_T} = \frac{D_T}{D_A},
\ee
which we can combine with the above to get
\be
D_T^2 = D_A^2  \sum_{j}W_{j 1}d_j ,
\ee
where $D_T$ is the quantum dimension of the phase including confined particles.
In particular, when we have $D_A=1$,
\be
D_T^2 =   \sum_{j}W_{j 1}d_j .
\ee

For more general values of the indices $i,b$, there is still some simplification we can do
\be
\frac{1}{D_A}\sum_{c, a}W_{i a} N^{c}_{ab} \theta_c d_c
= \frac{1}{D_B}\sum_{j,k} N^{k}_{ij} W_{j b} \theta_k d_k,
\ee
using the fact that the only non-vanishing contribution has $\theta_a= \theta_i$ and $\theta_j= \theta_b$, which can be taken out of the sum and be canceled between the two sides.
A challenge is to derive this from anyon condensation.

There are also various inequalities satisfied by the $W$ matrix, if we make use of the fact
that  $D_B> D_A$ and there is conservation of quantum dimension when we split anyons, which are representations of phase $B$ into anyons of phase $A$. 

Since we have conservation of quantum dimensions, we expect that in the decomposition
\be
i \to a\oplus b\oplus \cdots, \qquad d_i = W_{ia} d_a + W_{ib} d_b + \cdots
\ee
Some of these anyons after the decomposition will be confined, and invisible to $W$.

Therefore, we must have 
\be
d_i \ge \sum_a W_{ia} d_a.
\ee

\subsection{Verlinde formula and commutativity}
It appears that at the level of the commutativity formula it already knows about Verlinde formula.

Let us first review the Verlinde formula and some of its implications:
\be
N_{ab}^c = \sum_x \frac{S_{ax}S_{bx}S_{\bar{c} x}}{S_{1x}}.
\ee
Now, using also the expression for $S$ in (\ref{eq:Smatrix1}), this give a non-linear
constraint on the fusion rules and relates also the fusion with quantum dimensions
and spins.

In particular, we focus on taking $a=c=1$. In this case, $N_{1b}^1 = \delta_{1b}$ holds for a
UMTC.
Substituting this into the Verlinde formula, we end up with
\be\label{eq:VerlindeReduced}
\delta_{1b}=\frac{1}{D} \sum_x d_x S_{xb}.
\ee

Note that the above relation is already very similar to the commutativity formula except we have
\be
W_{1b} = \sum_x W_{1x} S_{xb}.
\ee

Now consider however taking the above formula, multiply it by $d_b$ and sum over $b$ as well.
Then we have on the left hand side
\be
\sum_b W_{1b} d_b = D_B,
\ee
where again we take phase $B$ to have quantum dimension $D_B$ and condense to
the trivial phase $A$ with $D_A=1$. This equation is a special case of Eq. \eqref{eq:DB/DA}.
The right hand side however gives
\be
 \sum_x \sum_bW_{1x} S_{xb} d_b = \sum_x W_{1x} \delta_{x1} D_B = D_B,
\ee
where we have made use of the special case Verlinde formula \eqref{eq:VerlindeReduced} in the first equality and $W_{11}=1$
in the second equality.
Therefore, commutativity is implicitly requiring consistency with Verlinde formula,
even though a priori it is not obvious that we input this into the definition of things.

\end{appendix}

\bibliographystyle{apsrev}
\bibliography{StringNet}
\end{document}